\documentclass[aps,prb,twocolumn,tightenlines,superscriptaddress,amsmath,amssymb,longbibliography]{revtex4-1}
\usepackage{graphicx}
\usepackage{bm}
\usepackage{comment}
\usepackage{graphicx,ams math,epsfig,epstopdf}% Include figure files
\usepackage{dcolumn}% Align table columns on decimal point
\usepackage{bm}% bold math
\usepackage{hyperref}% add hypertext capabilities
\usepackage{tabularx}
\usepackage{multirow}  
\hypersetup{
	colorlinks = True,
	urlcolor=blue,
	linkcolor = blue,
	citecolor= blue
}
\usepackage{grffile}

 \newcommand{\be}{\begin{equation}}
 \newcommand{\ee}{\end{equation}}
\newcommand{\bea}{\begin{eqnarray}}
\newcommand{\eea}{\end{eqnarray}}
\newcommand{\ba}{\begin{eqnarray*}}
\newcommand{\ea}{\end{eqnarray*}}

\usepackage{multirow}

\usepackage{color}
\usepackage[english]{babel}
\usepackage{dcolumn}% Align table columns on decimal point
\begin{document}

\title{Quantum Monte Carlo Annealing with Multi-Spin Dynamics}

\author{Guglielmo Mazzola}
 \affiliation{Theoretische Physik, ETH Zurich, 8093 Zurich, Switzerland}
\author{Matthias Troyer}
 \affiliation{Theoretische Physik, ETH Zurich, 8093 Zurich, Switzerland}
\affiliation{Quantum Architectures and Computation Group, Microsoft Research, Redmond, WA 98052, USA}

%\date{\today}
             
%%%%%%%%%%%%%%%%%%%%%%%%%%%%%%%%%%%%%%%%%%%%%%%%%%%%%%%%%%%%%%%%%%%%%
\begin{abstract}
We introduce a novel Simulated Quantum Annealing (SQA) algorithm which employs a multispin quantum fluctuation operator.
At variance with the usual transverse field, short-range two-spin flip interactions are  included in the driver Hamiltonian.
A Quantum Monte Carlo algorithm, capable of efficiently simulating  large disordered systems, is described and tested.
A first application to SQA, on a random square lattice Ising spin glass reveals that the multi-spin driver Hamiltonian improves upon the usual transverse field.
This work paves the way for more systematic investigations using  multi-spin quantum fluctuations on a broader range of problems.
 
\end{abstract}
%%%%%%%%%%%%%%%%%%%%%%%%%%%%%%%%%%%%%%%%%%%%%%%%%%%%%%%%%%%%%%%%%%%%%
  \maketitle
  
\section{Introduction}

Discrete combinatorial optimization problems can be encoded into minimizing the energy of classical Ising-type Hamiltonians\cite{fu1986application} and
it has been proposed that adding quantum mechanics could help in finding the ground state of these frustrated Ising system faster than any classical technique. One can define a suitable time-dependent quantum Hamiltonian that connects an initial Hamiltonian, whose ground state is easy to prepare, to the  final classical Ising Hamiltonian. At sufficient low temperatures a quantum system undergoing such adiabatic relaxation  reaches its ground state and solves the optimization problem. This technique is called quantum annealing (QA)\cite{farhi2000quantum,PhysRevE.58.5355,Farhi20042001,RevModPhys.80.1061}.

Early numerical studies predicted QA~\cite{PhysRevE.58.5355,Farhi20042001,Santoro29032002}  to be a competitive computational resource for solving spin glass problems,  compared to its closely related classical counterpart, the simulated annealing  (SA) algorithm~\cite{Kirkpatrick13051983}. However, so far no quantum speedup has been observed in experiments \cite{boixo2014,Ronnow:2014fd}.
 The essential difference between these two heuristic methods relies on the type of fluctuations which  drive the system away from the multiple local minima, occuring in rugged energy landscapes.
 
 Quantum fluctuations are expected to give an advantage to quantum annealing  in particular when the free energy landscape displays tall but narrow barriers.  These are easier to tunnel through quantum-mechanically, compared to climbing over them by means of thermally activated escape events.

In QA the system closely follows the ground state of a time-dependent Hamiltonian $H(t)$  which at $t=0$ is dominated by a pure quantum fluctuation part $H_Q$ whereas the final Hamiltonian $H(t_{\rm final})$ encodes only the cost function $H_P$ of the combinatorial optimization problem.  The Hamiltonian as a function of the time $t$ may read,  in the case of simple linear annealing schedules,
\begin{equation}
\label{e:qa}
H(t) = H_P + (t_{\rm final} - t)~H_Q~.
\end{equation}

Since random ensembles of hard problems are closely connected to spin glass models, we choose, for our tests,  the problem Hamiltonian as an Ising spin glass
\begin{equation}
H_P =  \sum_{i,j}  J_{ij} \sigma_i^z \sigma_j^z~,
\end{equation}
where $\sigma_i^{z}$ are Pauli matrices, acting on spins $i$, and $J_{ij}$ is the coupling between spins $i$ and $j$, which are randomly uniformly distributed in a range $[-1,1]$ .

Most QA studies so far employed a single type of the quantum term $H_Q$, a transverse-field (TF)
\begin{equation}
\label{e:tf}
H_Q^{TF} =- \Gamma \sum_i \sigma_i^x~,
\end{equation}
where the $\sigma_i^x$ operator acts locally on the spin index $i$ inducing quantum fluctuations. This is simplest to implement both in physical devices, such as the D-Wave devices\cite{johnson2011quantum,PhysRevB.82.024511,dwave2x,boixo2014}
 and in simulated quantum annealing (SQA) by means of quantum Monte Carlo (QMC) methods\cite{Santoro29032002,Heim12032015,isakov2015understanding,denchev2015computational}.

However, as pointed out by Ref.~\onlinecite{PhysRevE.75.051112}, it is possible to choose also different types of quantum fluctuations operators $H_Q$.
Following Ref.~\onlinecite{PhysRevE.75.051112} we define a two body fluctuation  operator with ferromagnetic interactions (FI)  as
 \begin{equation}
\label{e:hq}
H_Q^{FI} = - \Gamma \sum_i \sigma_i^x  -  \Lambda \sum_{i,j}  \sigma_i^x \sigma_j^x~,
 \end{equation}
where \emph{i,j} are nearest neighbours, so that the interaction is short-range. 

The adiabatic theorem states that the system follows the instantaneous ground state as long as the total annealing time $t_{final} \gg 1/\Delta^2$  where $\Delta$ is the minimum energy gap from the ground state, which is encountered along the annealing run of Eq.~(\ref{e:qa}).
It has been conjectured and shown for simple models, that employing quantum fluctuations beyond the TF term could be beneficial to avoid these small gap events
\cite{PhysRevE.75.051112, seonae2012}, which are the bottlenecks of QA\cite{Santoro29032002,Altshuler13072010,Farhi2012,knysh2015}.
Another argument in support of this possibility is that, in Eq.~(\ref{e:hq}), we are effectively adding a new schedule parameter $\Lambda$ which can be also optimized to increase the QA performance.

Moreover, it has been found that  quantum annealing with transverse fails to identify all degenerate ground-state configurations, preventing a fair sampling of equally probable states\cite{matsuda,mandra2016exponentially}.
Multi-spin quantum fluctuations could alleviate or possibly remove this issue.

In this paper we devise a QMC algorithm to perform SQA with this two-spin transverse ferromagnetic interaction. We note that the \emph{minus} sign in front of  $\Lambda$ makes this Hamiltonian \emph{stoquastic} for  $\Lambda>0$. This means that it is sign-problem free and can thus be simulated by  QMC methods.
On the other hand, the existence of a sign problem precludes the possibility to simulate the two spin transverse antiferromagnetic ($\Lambda<0$) interaction\cite{PhysRevE.85.051112,hormozi2016non,nishimori2016exponential} within the same approach unless the graph of couplings is bipartite.

The paper is organized as follows, in Sect.~\ref{s:qmc} we	describe the QMC algorithm, which is tested against exact results for small systems in Sect.~\ref{s:test}.
In Sect.~\ref{s:sqa} we report a first illustrative application to the relevant problem of Ising glass SQA.
We discuss the results as well as possible future developments in Sect.~\ref{s:conc}.

\section{Quantum Monte Carlo Methods}
\label{s:qmc}
\subsection{Overview}

Quantum Monte Carlo (QMC) methods are the only classical approaches for simulating quantum annealing on systems as large as the ones  realized experimentally, which consider more than $N=1000$ qubits.
Other numerical methods such as \emph{unitary evolution}\cite{PhysRevE.58.5355,Farhi20042001} scale exponentially  with $N$ and  are  therefore limited to much smaller numbers of spins ($N \approx 20$).

The most widely used QMC technique in the context of QA is \emph{Path Integral Monte Carlo (PIMC)}.
PIMC relies on the path integral formalism of quantum mechanics and samples the density matrix corresponding to the quantum Hamiltonian $H$ by means of a classical Hamiltonian $H_{cl}$ on an extended system having an additional dimension, the \emph{imaginary time} direction\cite{Santoro29032002,Heim12032015,isakov2015understanding}.
The original transverse field quantum spin system is then mapped into a classical one, which can be simulated by standard Metropolis Monte Carlo.

It has recently been shown \cite{isakov2015understanding,jiang2016scaling}  that for tunneling through a barrier the PIMC efficiency scales as a function of the system size as a physical QA  would .
This connection exists due to the fact that the tunneling rate in QMC and in the exact real-time quantum evolution scales in the same way, to leading order, as a function of the relevant parameters.  

This connection has  two implications. 
The first is that a QMC annealing simulation is relevant for investigating the behaviour of a real quantum annealer, as long as it is stoquastic and the QMC simulations thus do not suffer from a sign problem. 
The second is that it makes QMC-SQA a competitive tool compared to QA in real devices, by precluding a scaling advantage of stoquastic QA over SQA (cfn. also Ref.~\onlinecite{denchev2015computational}).

Starting from this mapping, other \emph{quantum inspired} algorithms can be developed such as PIGS, where open boundary conditions in imaginary time are used\cite{isakov2015understanding} for further acceleration. In this case QMC represents only a classical optimization algorithm and the physical meaning of the simulation is lost.
Similarly, it has been recently shown that performing PIMC away from the converged physical limit, provides a more efficient  algorithm  compared to the continuos time one\cite{Heim12032015}. This can be also considered a \emph{quantum inspired} algorithm.
For this reason we focus here on implementing a discrete-time PIMC algorithm rather than a continuos time PIMC.

\subsection{Local versus Cluster Updates}
\label{s:clustt}

The simplest Metropolis algorithm performs \emph{local} updates: one generates trial configurations by simply flipping one spin at a time.
%It is clear that, if $N_l$ is the lattice size of the system, a new uncorrelated configuration is sampled after $N_l$ accepted steps. {\color{red} \bf [This statement is actually wrong. We can be trapped for exponentially long times.]}
While this algorithm is ergodic, it can be very inefficient. For non-frustrated systems considerable improvements is obtained by cluster algorithms\cite{PhysRevLett.58.86,PhysRevLett.62.361}. They allow simultaneous flips of clusters
of spins and  are especially advantegeous  when large
scale fluctuations are important, usually around phase transitions, where local update algorithms slow down.

Standard cluster algorithms\cite{PhysRevLett.58.86,PhysRevLett.62.361} allow the clusters to grow without restrictions. While efficient for non-frustrated systems this approach breaks down for frustrated spin systems, as spin glasses. Due to the frustration the clusters grow to fill a very large fraction  of the lattice, if not the entire lattice. The clusters are then  larger than the physical domains that should be flipped and the algorithm becomes inefficient\cite{PhysRevB.45.4700,PhysRevB.43.8539,coddington1994generalized,PhysRevLett.115.077201}. Essentially the whole system freezes out and one flips almost all spins.

Therefore, in the context of SQA of spin glasses with transverse field an hybrid approach is usually employed, in which \emph{restricted} clusters are built\cite{rieger1999application,Heim12032015}.  Here, local clusters can grow on a single site only in the imaginary time dimension, where there is no frustration. In this paper we employ the same strategy with an important modification. Due to the non local nature of the two-spin interaction as cannot restrict ourselves to strictly local clusters but allow a small extent in the spatial direction.

\subsection{Cluster QMC for Pure Two-Spin Couplings}

\label{s:loop}

We start with the case $\Gamma=0$, i.e. considering only the  two-spin term  in Eq.~(\ref{e:hq}).
 The Hamiltonian $H = H_P + (t_{\rm final} - t)~H_Q$, at any time $t$ is as a special case of the anisotropic quantum XYZ model \cite{takahashi1972one}.
In this Section we present a modified version of  the \emph{loop algorithm} \cite{evertzloop} used to simulate the thermodynamics of this model. Readers unfamiliar with the loop algorithm arte referred to the review \cite{evertzloop} for details. 

For sake of simplicity we consider a ferromagnetic Hamiltonian with homogeneous couplings $J>0$ in this discussion, defined on an arbitrary graph of $N$ sites connected by $B$ bonds (the disordered case $J_{ij}$ will be considered at the end):
\begin{equation}
H = - \left(  J \sum_{i,j}   \sigma_i^z \sigma_j^z +  \Lambda \sum_{i,j}   \sigma_i^x \sigma_j^x \right)~.
\end{equation}
The generalization of the algorithm to spatially  varying couplings is straightforward.

\begin{figure}[ht]
\centering
\includegraphics[width=1\columnwidth]{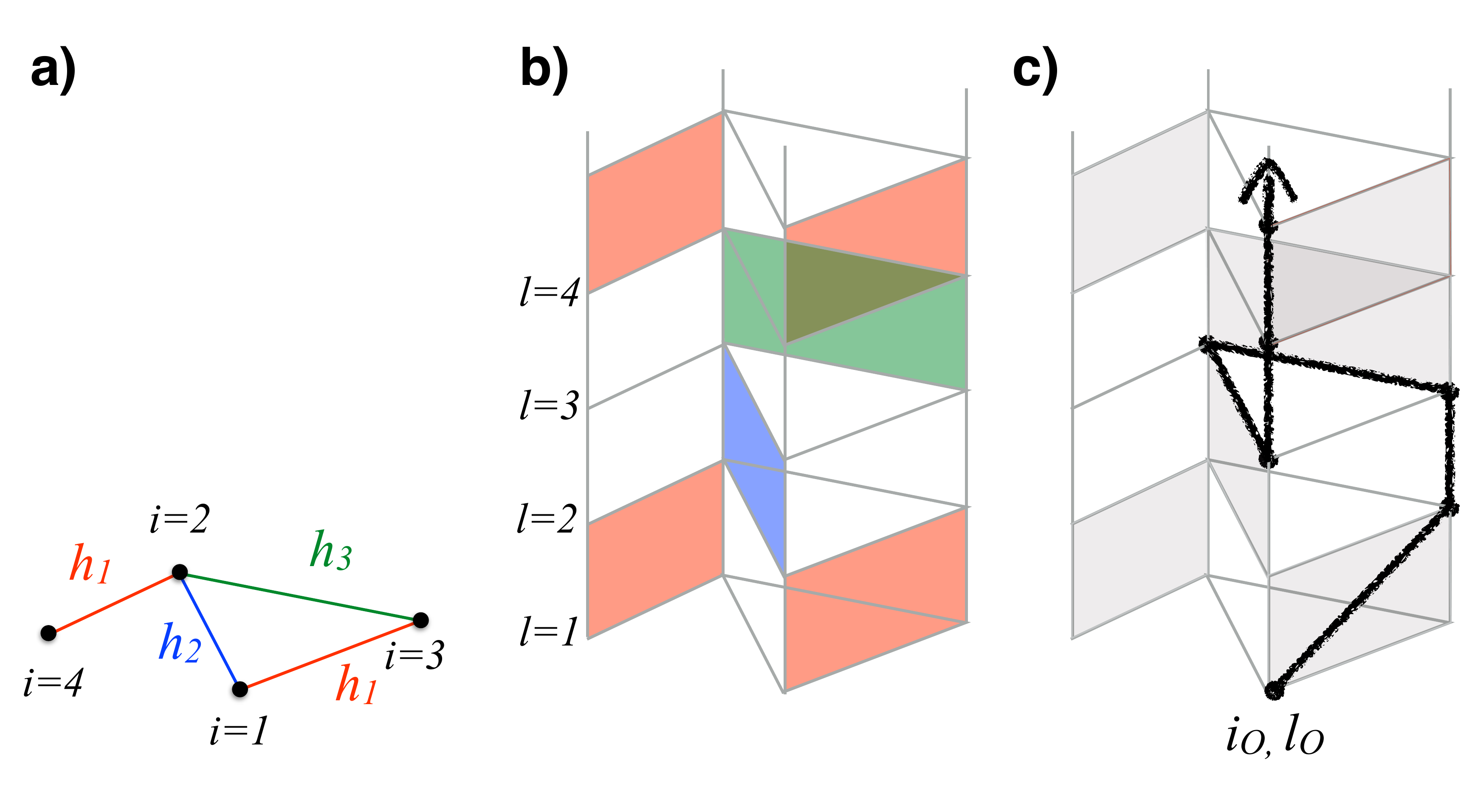}
   \caption{\emph{a)} Splitting of non-commuting bond Hamiltonians $h_k$ for a simple, non regular  graph with four sites. In this specific case   $K=3$. 
   \emph{b)} Extendend graph with shaded plaquettes construction along the imaginary time direction. The bond Hamiltonian $h_1$ acts at the time slice $l = 1~ \mathrm{mod} ~K$, etc. The primitive cell  repeats vertically each $K=3$ slices.
   \emph{c)} Example of cluster building (see Sect.~\ref{ss:algo}). In this sketch we add 6 sites after the initial $(i_O,l_O)$, following the \emph{Loop Algorithm} procedure detailed in Ref.~\onlinecite{evertzloop}. The cluster building will continue until we reach again the starting site $(i_O,l_O)$. 
       \label{fig:graf}}
\end{figure}

We first identify the non-commuting pieces of this Hamiltonian to perform a Trotter breakup. This procedure is less straightforward compared to the transverse field case where the Hamiltonian splitting was a trivial splitting $H=H_P+H_Q^{TF}$.
Here we need to split the Hamiltonian $H  = \sum_{k=1}^K h_k$ into a sum of commuting bond Hamiltonians $h_k$
\begin{equation}
h_k = \sum_{b \in \{b\}_k} H_{i_b,j_b} = \sum_{b \in \{b\}_k} H_b~ ,
\end{equation}
where $i_b,j_b$ are two sites connected by the bond $b$ and $H_b=-( J\sigma_{i_b}^z \sigma_{j_b}^z  + \Lambda    \sigma_{i_b}^x \sigma_{j_b}^x).$ Each set   $\{b\}_k $ is  defined such that its elements  don't share any site, i.e. no site $i$  appears twice (see Fig.~\ref{fig:graf}.a) .
The number $K$ of non-commuting  terms $h_k$ depends on the graph's connectivity. For example, in a linear chain we have $K=2$ and the standard checkerboard decomposition\cite{evertzloop}, whereas for a square lattice $K=4$.

The partition function then reads
\begin{equation}
Z = tr ~e^{-\beta H} = \lim_{M\rightarrow \infty} tr ~ \left( \prod_{k=1}^K e^{-{\beta \over M} h_k}  \right)^M .
\end{equation}
Inserting complete sets of $\sigma^z$ eigenstates, and following Ref.~\onlinecite{evertzloop} we write
\begin{equation}
\label{e:Z}
Z = \sum_{\{s_{il}^z\}} W(\{s_{il}^z\}) = \sum_{\{s_{il}^z\}} \prod_p W_p(\{s_p\})  ~,
\end{equation}
where the outer summation is carried over all possible spin configuration on the extended lattice. The index $i$ runs over the original graph sites $i=1,\cdots,N$, whereas $l$ label the imaginary time coordinate $l=1,\cdots,MK$.
The index $p$ extends over all the \emph{shaded} plaquettes of the extended lattice (cfn. Ref.~\onlinecite{evertzloop}).
We define a plaquette as the 4-spin configuration $p=p(b,l)=\{ (i_b,l),(j_b,l),(i_b,l+1),(j_b,l+1)$.
The statistical weight of each plaquette $p=p(b,l)$ is
\begin{equation}
W_p(s_p) = \langle  s(i_b,l) s(j_b,l) | e^{-\Delta H} | s(i_b,l+1) s(j_b,l+1) \rangle~,
\end{equation}
where $\Delta = \beta/M$. $W_p(s_p)$ is different from $1$ only when the piece of Hamiltonian $h_k$ acts on the bond $b$ at the 
correct time $l=(\ell-1)K + k$, with $\ell = 1,\cdots,M$.
This condition defines the \emph{shaded} plaquettes. 
For each bond $b$ we have exactly $M$ shaded plaquettes, one for each Trotter time step $\ell$ (see Fig.~\ref{fig:graf}.b) . 
The non-trivial Hamiltonian evolution in imaginary time occurs only on these special 4-spins configurations.
For the simple case of the spin chain, the shaded plaquettes construction resembles a checkerboard lattice.
There are only  \emph{eight}  non vanishing matrix elements on the shaded plaquettes, which can be divided in \emph{four} types:
\begin{align}
\label{eq:plaquette}
W(1) &=\langle ++ | e^{-\Delta H_b} | ++ \rangle \simeq 1+ \Delta J \\
W(2) &=\langle +- | e^{-\Delta H_b} | +- \rangle \simeq 1- \Delta J \\
W(3) &=\langle +- | e^{-\Delta H_b} | -+ \rangle \simeq  \Delta \Lambda \\
W(4) &=\langle ++ | e^{-\Delta H_b} | -- \rangle \simeq  \Delta \Lambda ~,
\end{align}
expanding the exponential for small $\Delta$, and taking into account that, for example $W(1) =\langle ++ | e^{-\Delta H_b} | ++ \rangle  =\langle -- | e^{-\Delta H_b} | -- \rangle$ since the one body magnetic field operator $\sigma_z$ is absent.
Since  $H_b$ preserves the parity of the magnetization, only 8 out of the 16 possible states give a non-zero contribution. % (see Fig.~\ref{fig:brks}).
Notice that this property would not hold in the transverse field case. 

The simplest Monte Carlo algorithm usually employs local updates, i.e. by proposing to flip one spin at a time. 
This is not a  good choice for XX couplings as unconstrained single spin flips  lead to forbidden plaquette configurations with zero weight. 
Our strategy is therefore to implement directly a cluster MC algorithm which automatically avoids sampling of forbidden configurations and minimizes the autocorrelation times.
The key feature of the \emph{loop algorithm}\cite{evertzloop} is that it allows \emph{nonlocal} changes of spins configurations, by  making  only \emph{local} stochastic decisions on the shaded plaquettes.
We refer the reader to the original Ref.~\onlinecite{evertzloop} for an exhaustive and rigorous justification of the algorithm.
Here we simply describe the outline of our optimized version for XX couplings, which, as it usually happens in most MC algorithms, can be divided in two steps: cluster formation and cluster flip.

\subsubsection{Outline of the Algorithm}
\label{ss:algo}

\begin{figure}[ht]
\centering
\includegraphics[width=1\columnwidth]{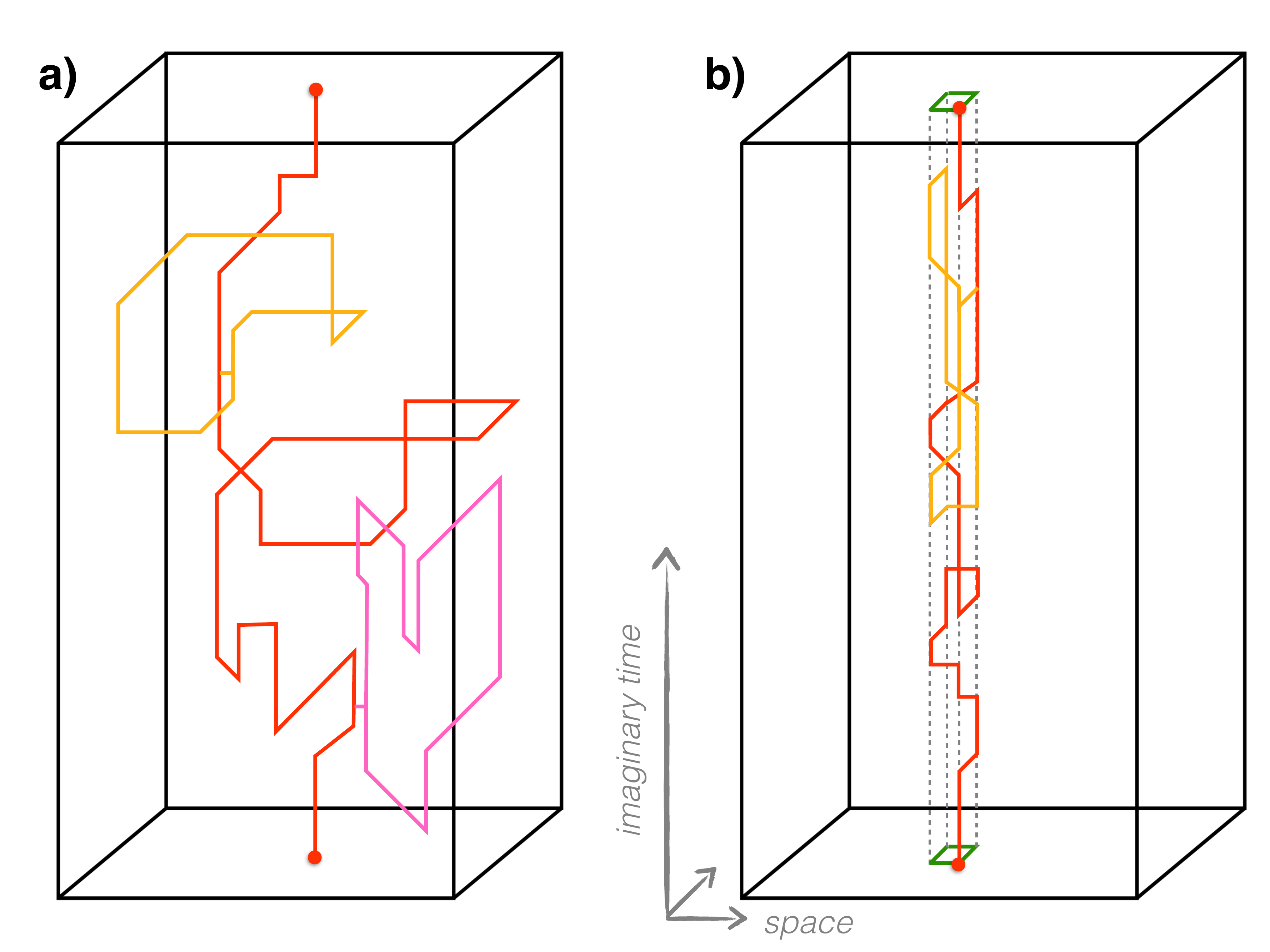}
   \caption{    Different types of cluster update, unrestricted in panel \emph{a)} and restricted, 
   \emph{b)} . Colors (red, orange and pink) depict loops that are glued together after that the cluster formation procedure is terminated, if  freezing breakups occur (cfn. Ref~\onlinecite{evertzloop}).
   In the restricted case, the cluster is confined into a 4 bond loop in the physical lattice (green square), other choices are obviously possible. 
   While in the unrestricted case the cluster is always flipped, in the restricted case the cluster is flipped according to  a suitable Boltzmann probability (see text).
   \label{fig:grafo}}
\end{figure}

We first restrict the growth of the cluster to a subset of bonds $\mathcal{B}_m$ such that $\bigcup_m \mathcal{B}_m = \mathcal{B} $ contains all the $B$ bonds of the graph (see Fig.~\ref{fig:grafo}). 
For example, a possible choice could be these set of bonds which constitute all the possible  smallest loops in the graph, as in Sect.~\ref{s:resu}.
This  is necessary to keep the cluster size small in the $N \rightarrow \infty$ limit (see Sect.~\ref{s:clustt}). 

 Next we explicitely show how to combine to standard \emph{Loop algorithm}\cite{evertzloop} with the restricted update scheme, therefore in the following we implicitely refer to the \emph{Loop algorithm} terminology and we suggest that the reader should  become familiar first with the standard version described in Ref.~\onlinecite{evertzloop}.

The algorithm proceeds as follows:

\emph{i.} We randomly select the set $\mathcal{B}_m$ of spins on which we want to build a cluster. Then we choose one bond $b_O \in \mathcal{B}_m$ and a Trotter time-slice $\ell = 1,\cdots,M$. 
This identifies the corresponding shaded plaquette $p=p(b_O,l_O)$.
 Finally we  randomly select one of the two sites, i.e. $(i_O,l_O)$, connected by the bond $b_O$, at the imaginary time $l_O$. This will be the first site added to the cluster 

\emph{ii.} We follow the \emph{Loop Algorithm} rules to add spins to the cluster. In this case we  add only spins which are connected by a bond in the set 
$\mathcal{B}_m$. 

\emph{iii.} We propose to flip this cluster with suitable probability $p_{flip}$, i.e. we reverse each spin $s \rightarrow -s$  belonging to the cluster.
$p_{flip} = \mathrm{min}[1,\mathrm{exp}(\delta E ~K/\Delta)]$ is given by the change in energy $\delta E$ due to the cluster flip.
Notice that only the bonds outside the set $\mathcal{B}_m$ will be taken into account in computing $\delta E$ since the cluster building rules take care of detailed balance for all bonds within $\mathcal{B}_m$. In this way we evaluate the energy gain of the cluster flip  restricted to $\mathcal{B}_m$, which is embedded in the graph. This is the main difference compared to Ref.~\onlinecite{evertzloop}.

Notice also that if $\mathcal{B}_m = \mathcal{B}$ already contains all the possible edges of the graph, then step \emph{iii.} simplifies to flipping the cluster with probability $1$, and we recover the single cluster formulation of the Loop Algorithm. We label this type of move as \emph{unrestricted} or \emph{global} update.
If instead the cluster is restricted to be on a localized set of bonds, we are performing a \emph{semi-local} update.

\subsubsection{Breakup Probabilities}
\label{s:brk}

The essential ingredients of the \emph{Loop Algorithm} are the so-called ``breakups weights".
In short, this set of   weights $w_{ij}$ ($i,j=1,2,3,4$) determines the shape and the size of the clusters.
For example, the weight $w_{ij}$ sets the probability that a shaded plaquette of type $i$ is changed into a plaquette type $j$, after the accepted MC update [cfn. Eq.~(\ref{eq:plaquette})].
In particular, this probability is given by $p_{i\rightarrow j} = w_{ij}/W(i)$.
For example, if we connect and flip the spins \emph{at the bottom} of a type-1 plaquette, we obtain a type-4 plaquette.
The flip of two spins inside a shaded plaquette always produce a different type of configuration, which is in turn a valid plaquette configuration, by construction. This is the main idea of the \emph{Loop algorithm}.
One needs to define every possible $i\rightarrow j$ transition weight. Some of these weights can also be zero, this means that some particular transitions are not-allowed.

We refer the interested reader to the original \emph{Loop Algorithm}\cite{evertzloop} for details, derivations and discussions. Here we simply provide our choice for the breakups weights for future reproducibility.

While the plaquette weights $W(i)$, with $i=1,2,3,4$, are fixed by the Hamiltonian [cfn. Eq.~(\ref{eq:plaquette})], there is freedom in choosing the breakup weights $w_{ij}$, provided that the following constraints are met
\begin{align}
\sum_j w_{ij} &= W(i), \label{e:constr}\\
 w_{ij} &> 0~.
\end{align}
Experience showed that one should optimize these weights for an efficient algorithm. 
Indeed, the autocorrelation time dramatically increases with non-optimal choices of $w_{ii}$, which gives the probability to remain in the same plaquette configuration. We thus aim to  minimize this weight.
The optimal weight's set varies as a function of the annealing schedule. Let us assume that $\Delta$ is small so that $W(1),W(2) > W(3),W(4) > 0$.

 Let us consider first the case $\Lambda > |J|$, i.e. at the beginning of the annealing.  
Under this condition  it is possible to set all the freezings to zero. The non-zero weights are:
 \begin{align}
  w_{12} &= 1 - \Delta J \\
  w_{13} &=w_{14} = \Delta J \\
  w_{34} &= \Delta(  \Lambda - J)~,
 \end{align}
 with $w_{ij}=w_{ji},~i,j=1,2,3,4$.
In the later stage of the annealing we have $\Lambda < |J|$ instead, so we find 
 \begin{align}
  w_{12} &= 1 - \Delta J \\
  w_{13} &=w_{14} = \Delta \Lambda \\
  w_{11} &= 2\Delta( J - \Lambda ) ~.
 \end{align}
 In this case is not possible to have $w_{ii} = 0$, but this is the optimal choice as pointed out also in Ref.~\onlinecite{kawashima1996cluster}.
 Notice that these two sets always satisfy Eq.~(\ref{e:constr}) and work both for ferromagnetic and anti-ferromagnetic couplings.

 \subsection{Adding the Transverse Field}
 \label{ss_tf}
 To obtain an algorithm for the full Hamiltonian
 \begin{equation}
 \label{e:fullham}
H =  \sum_{i,j}  J_{ij} \sigma_i^z \sigma_j^z -   \Gamma \sum_{i}   \sigma_i^x  -  \Lambda \sum_{i,j}   \sigma_i^x \sigma_j^x ~,
\end{equation}
we need to include
 the transverse field term $\sigma_i^x$ to the Loop Algorithm ($\Gamma=0$) described in Sect.~\ref{s:loop}. 
  This extension is hinted in Ref.~\onlinecite{evertzloop} and is not trivial as the transverse field term  $\sigma_i^x$  breaks the parity of the magnetization within each interaction plaquettes. 
  Instead of enlarging the total number of allowed plaquette configurations, taking into account all these possible states, we  treat this operator stochastically by adding additional single-site breakups.
The transverse field operator  can end a  loop cluster  at any spacetime point $(i,l)$, whereever $\sigma_i^x$  acts.
In our algorithm we add the possibility to stop the cluster when entering into a shaded plaquette at site $i$, with probability given by $p_x =\sinh{(\Delta \Gamma/ K_i)}$, where  $K_i$ denotes the number of physical neighbours of the site $i$ in the graph. This is the generalization  to arbitrary graphs of the procedure sketched in Ref.~\onlinecite{evertzloop} for a linear chain
In the following we give details on the actual implementation of this idea.

\emph{i.} Suppose we start the cluster from position $(i_0,l_0 )$.

\emph{ii.} When we jump into a shaded plaquette, say at position $(i,l)$, if the plaquette type is not of type 3 or 4\footnote{For simplicity we don't break clusters on these types of plaquettes. This approximation is justified in the $M\rightarrow\infty$ limit.},  we stop the cluster with probability $p_x$.

\emph{iii.}  Each time we stop we keep track of this position by inserting a $\sigma_i^x$ operator label. 
We  put this label on the bond above (below) site $i$,  if the direction was upward (downward) in imaginary time direction.
Otherwise we continue to build the cluster following the procedure described in Sect.\ref{ss:algo}.

\emph{iv.} If we stop the cluster then we restart from $(i_0,l_0 )$ and proceed in the opposite direction until we stop again, notice that now we can stop either by inserting a new $\sigma_i^x$ operator or by touching an existing one already in place. %(case \emph{3} of Fig.~\ref{fig:brksTF}).

\emph{v.} The cluster flipping decision remains unchanged. 

\emph{vi.} Finally, remove the $\sigma_i^x$ operator labels between segments of equal orientation. This completes one update.

The last issue concerns the existence of plaquettes having an odd number of spin up (down), which fall outside the breakup selection rule of the XX loop algorithm. These plaquettes certainly occur if $\Gamma>0$. Since in our approximation we can encounter the $\sigma_i^x$ operator only going vertically along the imaginary time direction, the decision rule can be adapted from the standard transverse field cluster algorithm\cite{rieger1999application} in this case:
suppose we are at site $(i,l)$ and we are proceeding upward, if there is not  a  $\sigma_i^x$ operator already in place above $(i,l)$, then we check whether the spin at $(i,l+1)$ is parallel to $(i,l)$, or not . In the latter case we stop the cluster. %(case \emph{4} of Fig.~

Finally we notice that, in the $\Lambda = 0$ case, this algorithm  does not reduce to the common TF algorithm for disordered systems\cite{rieger1999application}, in which clusters are built only along the imaginary time direction, if a non-empty bond set $\mathcal{B}_m$ is considered. Indeed, the cluster can still span the entire bond region $\mathcal{B}_m$, due to the occurence of the freezing plaquettes, which make the cluster non-local.
We use therefore the same cluster algorithm, within the same choice of  $\mathcal{B}_m$'s,  to compare FI and TF Hamiltonians, also considering the limiting $\Gamma = 0$ case. 
 
 \section{Phase Diagram of the XX Model with a Transverse Field}
 \label{s:test}
\begin{figure}[ht]
\centering
\includegraphics[width=1\columnwidth]{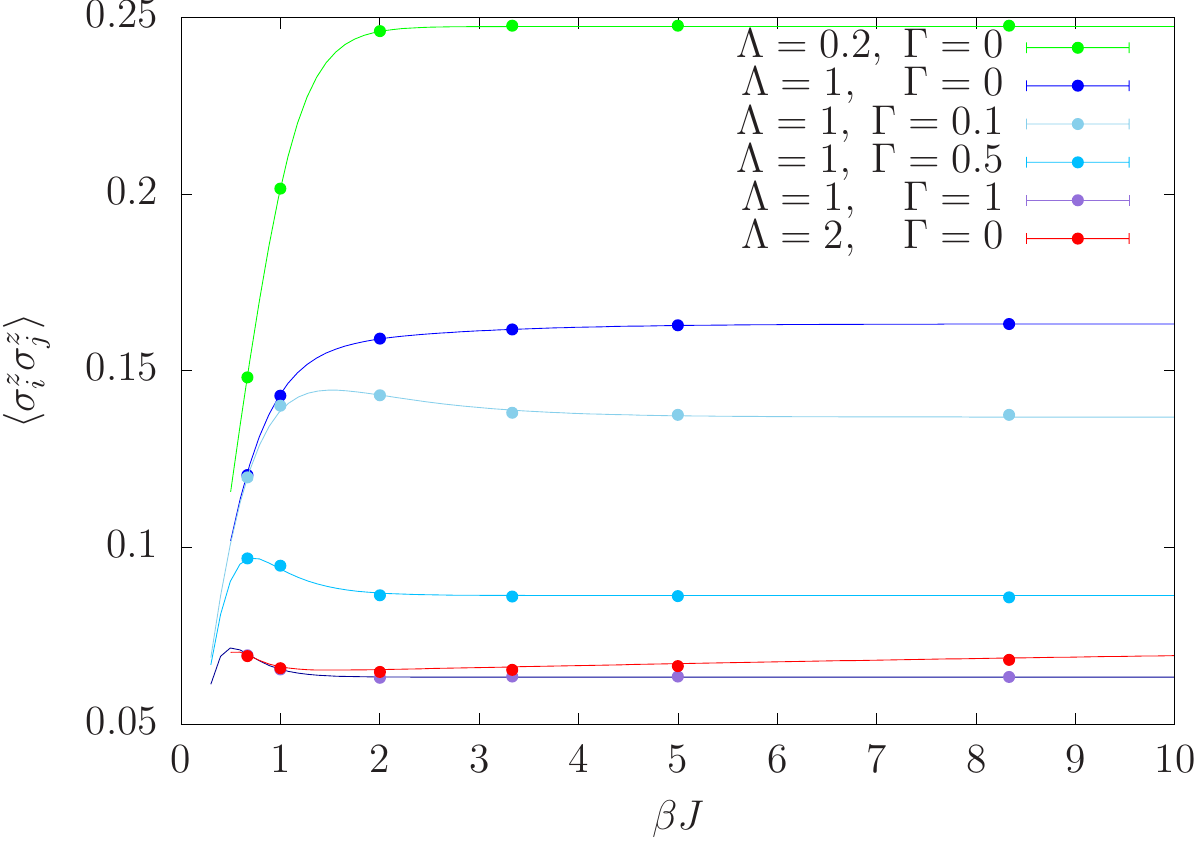}
   \caption{Nearest neighbours $\langle \sigma^z_i \sigma^z_j \rangle$ correlation function as a function of the inverse temperature $\beta$, for several choices of the parameters $\Lambda$ and $\Gamma$. 
The system is a ferromagnetic 8-site chain. 
   Points label QMC results, while continuos lines represent ED results. 
   For each temperature we use an appropriate number of Trotter slices $M$ to ensure convergence to the continuos imaginary time limit.
       \label{fig:corrvsbeta}}
\end{figure}

We test the accuracy of the algorithm against exact diagonalization (ED) results\cite{alps}, for a small uniform ferromagnetic chain and square lattices with spacial periodic boundary conditions, having Hamiltonian given by Eq.~(\ref{e:fullham}), i.e. without bond disorder.
In Fig.~\ref{fig:corrvsbeta} we plot the $\langle \sigma^z_i \sigma^z_j \rangle$ correlation function, for nearest neighbours, as a function of the inverse temperature $\beta$, for several choices of the parameters $\Lambda$ and $\Gamma$.
The agreement of QMC with ED is satisfactory in the full temperature range.

\begin{figure}[ht]
\centering
\includegraphics[width=1\columnwidth]{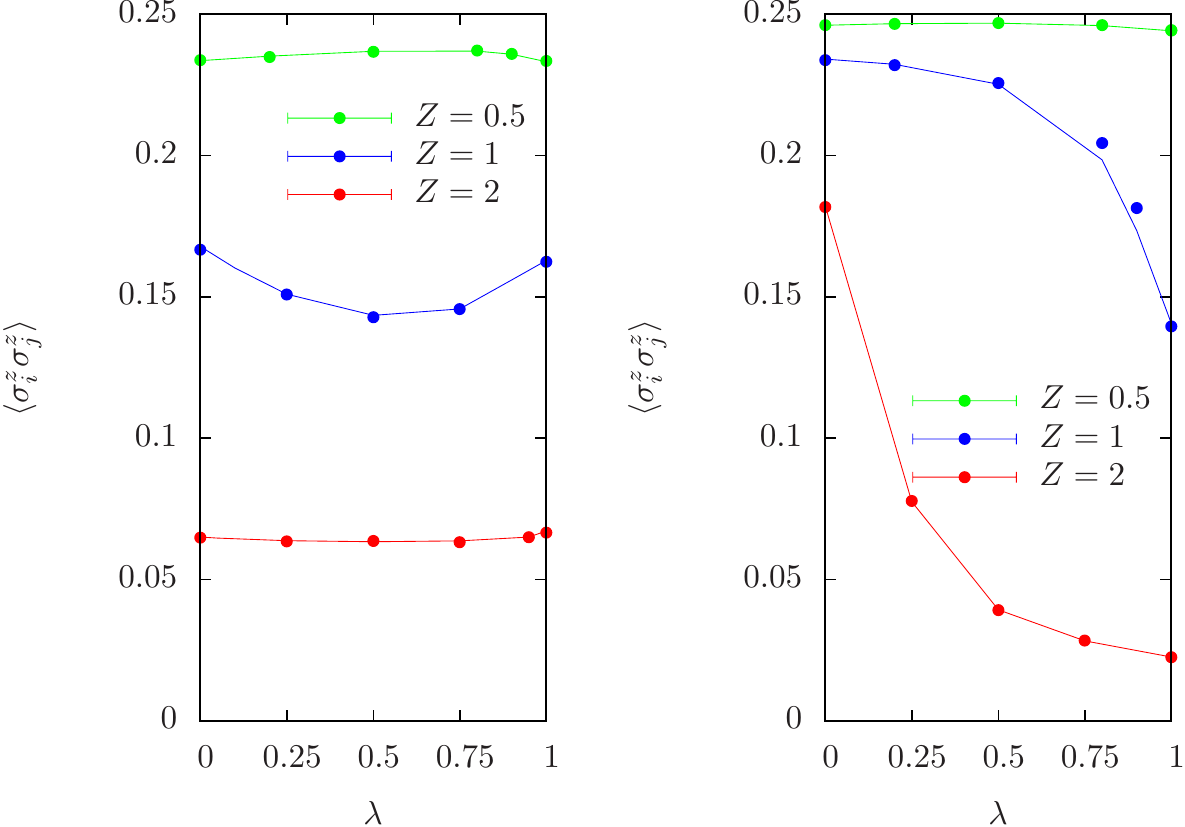}
   \caption{Low temperature ($\beta J=5$) nearest neighbours $\langle \sigma^z_i \sigma^z_j \rangle$ correlation function for several combinations of $\Lambda$ and $\Gamma$. We parametrize these values as $\Lambda = \lambda Z$ and $\Gamma = (1-\lambda) Z$ and we plot $\langle \sigma^z_i \sigma^z_j \rangle$ as a function of $\lambda$, for $Z=0.5,1,2$.
   Points label QMC results, while continuos lines represent ED results. 
   The left panel refers to a 8-sites ferromagnetic chain, while the right panel to a 16-sites square lattice.
       \label{fig:iso}}
\end{figure}

Next, in  Fig.~\ref{fig:iso}, we focus on the low-temperature regime relevant for SQA and benchmark simulations obtained with different combinations of $\Lambda$ and $\Gamma$ values, which can be realized along a generic SQA run.
 
\begin{figure}[ht]
\centering
\includegraphics[width=1\columnwidth]{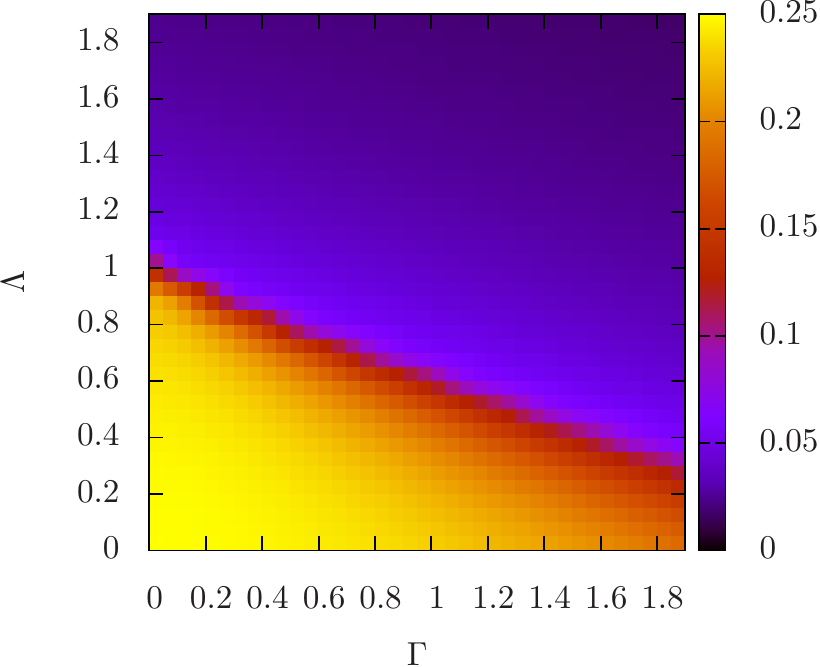}
   \caption{Low temperature ($\beta=20$) phase diagram of the model (\ref{e:fullham}), with $J_{ij}=1$,  on a $10\times10$ square lattice.
   We plot the $\langle \sigma^z_i \sigma^z_j \rangle$ correlation, with $i,j$ nearest neighbours, as a function of $\Lambda$ and $\Gamma$ parameters, both in the range $[0,1.9]$. 
   The phase boundary between the ordered (below) and disordered  (above) phases roughly lies  along the  $ \Lambda + \Gamma/2 =1$ line.
       \label{fig:phase}}
\end{figure}

Finally in Fig.~\ref{fig:phase}, we show, as an example, the low-temperature phase diagram of the Hamiltonian in Eq.~(\ref{e:fullham}), with $J_{ij}=1$, on a 2D square lattice.
Generalizing the breakup weights in Sect.~(\ref{s:brk})  for an arbitrary XYZ Hamiltonian, it would be possible to study the phase diagram of  sign-problem free XYZ models with transverse field on arbitrary lattice.

 \section{Simulated Quantum Annealing}
 \label{s:sqa}
 \subsection{Annealing schedule}
 
We now report a first application of the algorithm to SQA, to assess the annealing sensitivity to the $H^{FI}_Q$ driver Hamiltonian. 
We consider a spin glass  problem Hamiltonian, defined on a $10 \times 10$ square lattice, with periodic boundary conditions, and uniformly randomly distributed couplings in the range $[-1,1]$.
Following previous SQA studies\cite{Heim12032015,Santoro29032002}, we plot the residual energy $E_{res} (t_{final})= E(t_{final}) - E_0$ as a function of different annealing times $t_{final}$, where $E_0$ is the exact solution\cite{spinserver} and $E(t_{final})$ is the ground state of the Hamiltonian found at the end of the annealing.

In order to compare the performance of different annealing strategies in the $\Gamma-\Lambda$ parameters space, we need to rigorously define the computational effort for a QMC algorithm having unrestricted and semi-local updates.
Different annealing schedules lead to  different average cluster sizes, and, the larger is the cluster built, the heavier is the computational cost of each update. This cost is proportional to the cluster size $\bar{n}$ and to the number of bonds one has to check for evaluate $\delta E$ in the acceptance/rejection step. We notice that, in the algorithm, the computationally expensive operations are made on the shaded plaquettes.
Therefore, the total effort is still proportional to the number of sites $N$ (or generically to the bond number $B$) and the number of Trotter
 slices $M$, although the total number of slices along the imaginary time direction is $ M \times K$.
 
Here we thus define the computational cost associated we each annealing run  as
 \begin{equation}
 \label{e:effort}
\mathcal{C} (t_{final}) = t_{final} ~\bar{n}, 
 \end{equation}
where $t_{final}$ is the total number of Monte Carlo updates.

We notice that the QMC algorithm is efficient, for low $\Lambda$ and in the thermodynamic limit, only if we restrict the cluster growth, at each update, to a small bond subset with $\# \mathcal{B}_m \ll B $. In this way $\bar{n} \ll N \times MK$ 

Usually, with TF-SQA,  a linear schedule with starting field $\Gamma_0 = \Gamma(t=0) > \mathrm{max}~ |J_{ij}| $ is employed\cite{Heim12032015}, in our notations, this would correspond to decrease linearly the transverse field parameter as $\Gamma(t)= \Gamma_0 (1-t)/t_{final}$ and set $\Lambda=0$, $\forall t$.
In the following we will compare to the simplest annealing path for SQA with ferromagnetic interactions (FI-SQA), given by  
\begin{align}
\label{e:annpath}
\Gamma(t) &= \Gamma_0 (1-t)/t_{final} \nonumber \\
\Lambda(t)&= \Lambda_0 (1-t)/t_{final}~,
\end{align}
namely, decreasing linearly both the control parameters. 
We will also compare to the choice $\Lambda(t)= \Lambda_0 (1-t)/t_{final}$ and $\Gamma=0$, $\forall t$, i.e. employing only the two pure two-body operator. 
The optimization of the annealing path to non-trivial $\Lambda(t), \Gamma(t)$ time-dependences is left for future studies.

\subsection{ 2D Spin Glass Results}
\label{s:resu}

\begin{figure}[]
\centering
\includegraphics[width=1\columnwidth]{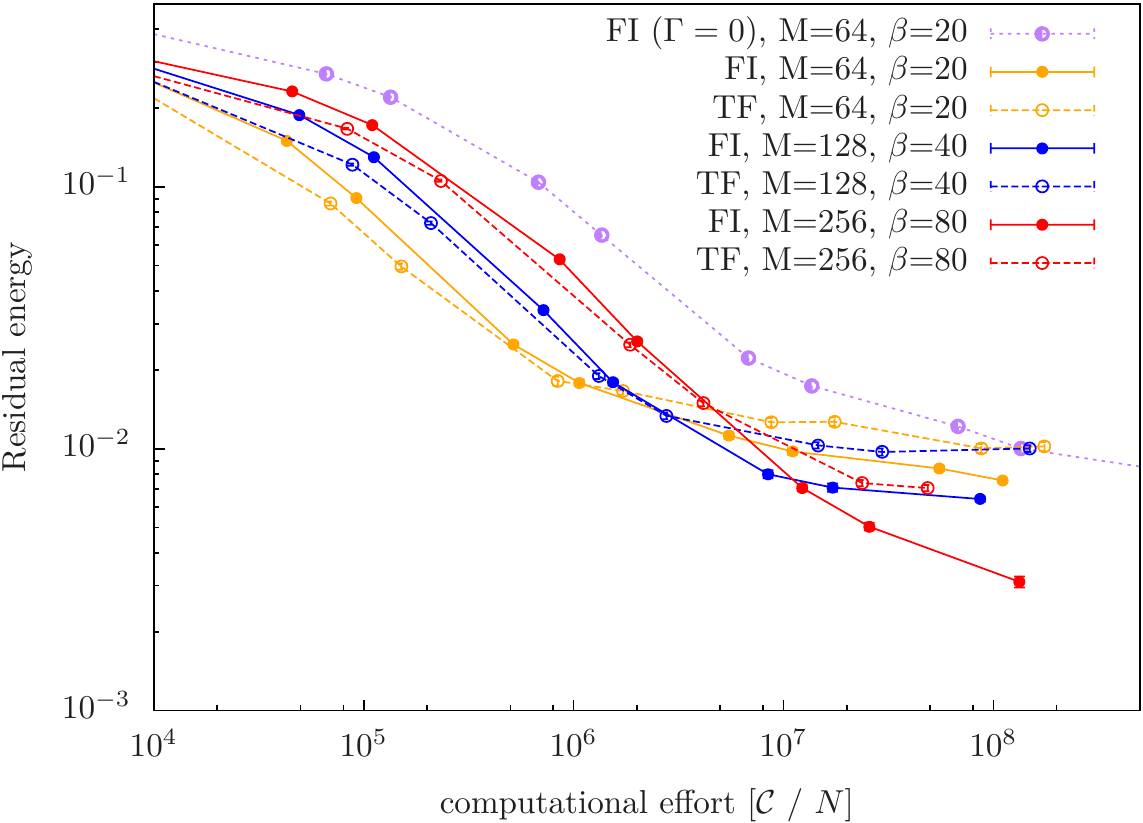}
    \caption{Residual energy $E(t_{final})$ as a function of total annealing time $t_{final}$, renormalized in order to express the total theoretical computational effort of each run as in Eq.~(\ref{e:effort}). 
Here, the QMC algorithm employs restricted cluster updates.    
    Dashed lines with empty symbols indicate SQA with standard TF Hamiltonian (with $\Gamma_0=2$), while continuos ones with solid symbols refer to FI-SQA, with annealing schedule given by Eq.~(\ref{e:annpath}), with $\Gamma_0=1$ and $\Lambda_0=1$.
    Each color (orange,blue, and red) labels different SQA setup, having diffent $M$ and $\beta$.  
    The FI Hamiltonian always performs better, for sufficiently long $t_{final}$, than the TF, within the same $M,\beta$ setup.
    Finally, we perform SQA without TF component (purple dot-dashed line). This choice gives poor perfomances.
       \label{fig:glass_mini}}
\end{figure}

In Fig.~\ref{fig:glass_mini} we show the residual energy as function of the annealing time, for  $10\times10$ lattice, and using the semi-local loop update, an approach that remains efficient for large disordered systems.
In this case, each possible bond subset $\mathcal{B}_m$, with $m=1,\cdots,N$, is defined to be the smallest 4-bonds loop that can be constructed on the square lattice (see Fig.~\ref{fig:grafo}).
The four lattice sites which belong to these sets are given by $(i,j),(i+1,j),(i+1,j+1),(i,j+1)$, with $i,j=1,\cdots,L$.

With this choice the QMC algorithm is ergodic as the original \emph{loop algorithm} (with global updates) and maintains efficiency in the disordered case, in the $N\rightarrow \infty$ limit.
In the general case, each bond subset $\mathcal{B}_m$ has to be provided as an input, and varies with the graph under consideration.

The bond subsets defining each non-commuting Hamiltonian $h_k,~ k=1,\cdots,K$ is also an input of the algorithm.
Generally, performing this decomposition can be also an hard optimization task, which is related to the \emph{edge coloring} problem, but it can always be solved by using a sub-optimal $K$, one larger than the minimal value (which depends on the graph complexity)\cite{edgecoloring}.
For regular graphs, such as the square lattice,  these sets arte always  easy to construct. 

In this study we compare TF and FI-SQA as \emph{classical} optimization algorithms. 
Therefore we renormalized the annealing time in such a way to fairly compare  the theoretical computational effort of each run, as discussed above.
 We also define the residual energy as the minimum energy $E_{min}$ among all the possible $M$ Trotter slices.
 We  perform the simulations at different  inverse temperatures $ \beta=20$, 40, and 80.
We keep the Trotter time-step $\beta/M = 0.3125$, constant and close to convergence to the physical limit (cfn. also Fig.~\ref{fig:glass_qa}).
 
In Fig.~\ref{fig:glass_mini} we observe that, for fixed $M$ and $\beta$, TF and  FI-SQA display different behaviour, as
 the two curves are not related by a trivial \emph{shift} along the time axis.
 We notice that the FI driver always outperforms the standard TF, for sufficiently large annealing times, at each temperature, despite the larger complexity of the FI driver.
Interestingly, the lower is temperature, the larger is the difference between the FI and TF residual energies.

Surprisingly, performing the annealing at zero transverse field, $\Gamma=0$, results in a drastic decrease of the performance.
These observations can be explained in the following way: 
\emph{i.} For short annealing time,  a large part of the residual energy can be recovered by simple one-flip moves, then the TF driver is more effective in eliminating the \emph{defects} in the extended lattice.
 \emph{ii.} The pure two body driver Hamiltonian $\sigma_i^x \sigma_j^x$ is way less efficient in this regards, as the clusters are usually bigger and always closed in the extended lattice.
 Therefore, the transverse-field operator is still an important ingredient for a QA driver Hamiltonian. 
\begin{figure}[]
\centering
\includegraphics[width=1\columnwidth]{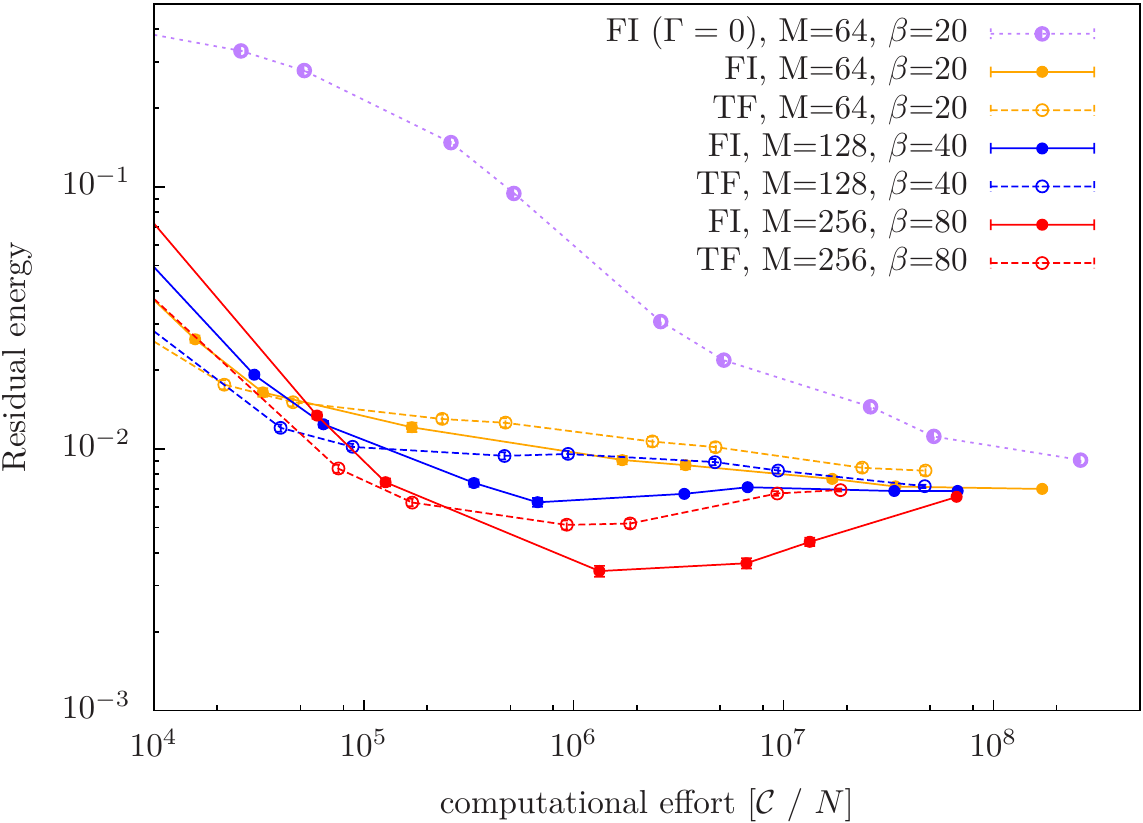}
   \caption{Residual energy $E(t_{final})$ as a function of total annealing time $t_{final}$, renormalized in order to express the total theoretical computational effort of each run as in Eq.~(\ref{e:effort}). 
Here, the QMC algorithm employs unrestricted cluster updates.    
    Dashed lines indicate SQA with standard TF Hamiltonian (with $\Gamma_0=2$), while continuos ones refer to FI-SQA, with annealing schedule given by Eq.~(\ref{e:annpath}), with (with $\Gamma_0=1$, $\Lambda_0=1$).
    Each color (orange,blue, and red) labels different SQA setup, having diffent $M$ and $\beta$.  
    The FI Hamiltonian always performs better, for sufficiently long $t_{final}$, than the TF, within the same $M,\beta$ setup.
       \label{fig:glass_loop}}
\end{figure} 
 
 Finally, we also check the performance using the global update algorithm. In this case the cluster can freely traverse the extended lattice.
We consider the same system and perform the same set of simulations and show the results in Fig.~\ref{fig:glass_loop}.
 We notice that the trend is preserved, although the annealing profiles are slightly different.
 In particular, the pure two-body driver performs significantly worse than the mixed FI-SQA, having both $\Lambda$ and $\Gamma$ non-zero.
 However, FI-SQA still outperforms TF-SQA for sufficiently long annealing times, as in the previous semi-local algorithm.
 Interestingly, this global updates version of the algorithm displays a better efficiency compared to the previous approach. 
 Indeed such system size can be considered still small and the global update scheme does not suffer much from a critical slowing down.
 
 Note though the  non-monotonicity as a function of the annealing time compared to the TF (cfn. also Ref.~\onlinecite{Heim12032015}), which hints at inefficiencies of the global updates. This feature disappears when the average energy is used (see Fig.~\ref{fig:glass_qa}), instead of the lowest one among the Trotter slices.

\section{Efficiency of Quantum Annealing}

\begin{figure}[]
\centering
\includegraphics[width=1\columnwidth]{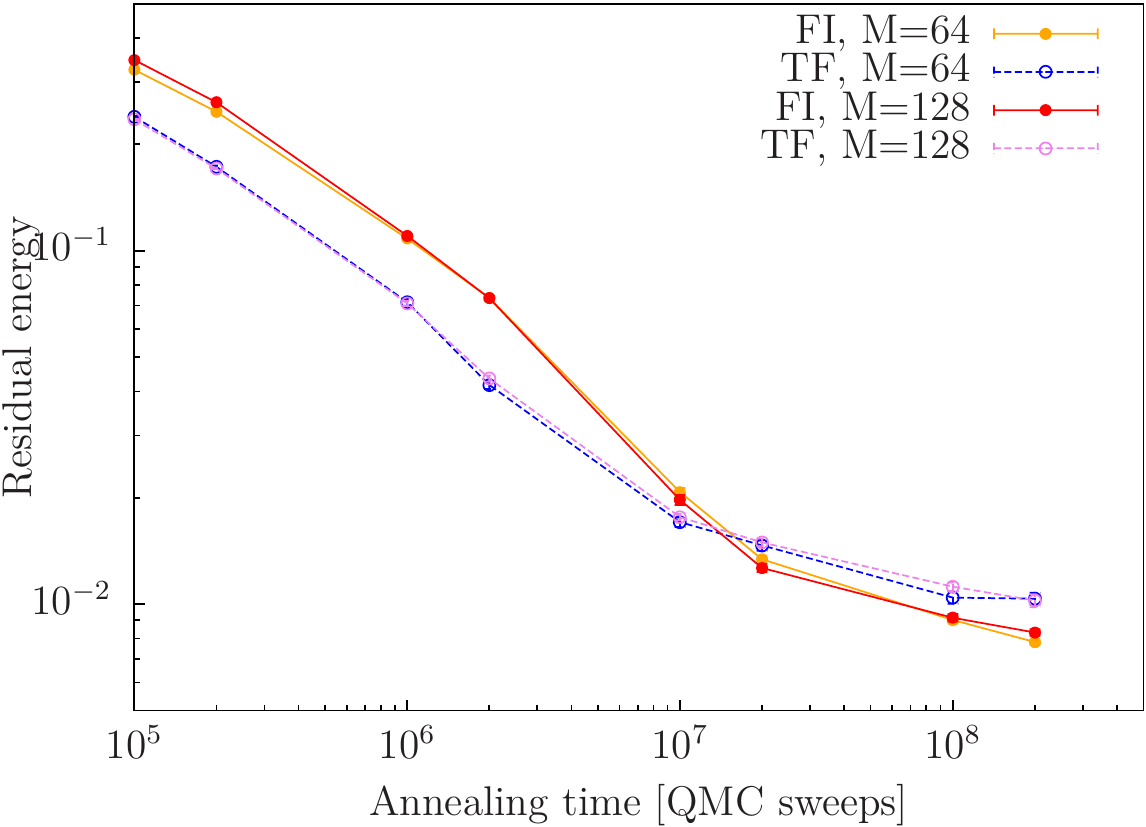}
   \caption{Residual energy $E(t_{final})$ as a function of total annealing time $t_{final}$ for different operators and Trotter slices $M$..
Here, the QMC algorithm employs the same restricted cluster updates for the different operators TF and FI.
The annealing time is  not renormalized taking into account the computational effort (which is proportional to $M$) and we obtain the final energy by averaging over all the Trotter slices. 
    Dashed lines indicate simulations with standard TF Hamiltonian (with $\Gamma_0=2$), while continuos ones refer to the FI operator, with annealing schedule given by Eq.~(\ref{e:annpath}), with (with $\Gamma_0=1$, $\Lambda_0=1$).
    We use an  inverse temperature $\beta=20$ and we check the continuos time limit convergence using $M=64$ and $128$.
    The FI Hamiltonian always performs better, for sufficiently long $t_{final}$, than the TF, within the same $M,\beta$ setup.
       \label{fig:glass_qa}}
\end{figure}

In this last section we  perform simulations indicative of the relative efficiency between the FI and TF operators in a real QA device instead of as a classical optimization algorithm. QMC is expected to reproduce the performance of physical QA  for typical tunneling problems as discussed in 
Refs.~\onlinecite{isakov2015understanding,jiang2016scaling}.
This time we take the continuous time limit by using  a large enough number of time slices $M$, and we measure the annealing time simply using the Monte Carlo steps, as the computational effort related to the different update scheme is not relevant for the study of the physical machine.
Moreover, following Ref.~\onlinecite{Heim12032015} we average the final energy over all the $M$ Trotter slices.
In Fig.~\ref{fig:glass_qa} we observe the same trend as in SQA, where the FI operator eventually outperforms the TF at later stages of the annealing.

\section{Conclusions}
\label{s:conc}
We introduced a novel QMC algorithm to perform SQA with a transverse ferromagnetic  two-spin driver Hamiltonian Eq.~(\ref{e:hq}).
This type of fluctuation extends the standard TF-SQA, adding also a two-body transverse operator of the form $\sigma_i^x \sigma_j^x$.
Though  this possibility has been introduced already some years ago\cite{PhysRevE.75.051112},  it has never been used before with QMC simulations, and therefore never applied so far on large optimization problems.
We notice that, since QMC techniques are limited by the so-called \emph{sign problem}, we can simulate only the ferromagnetic version of two body transverse interaction.

Our discrete-time path-integral Monte Carlo algorithm is an extension of the well established \emph{Loop Algorithm}\cite{evertzloop}, with the inclusion of the transverse field operator, and implements restricted cluster updates to simulate efficiently large disordered systems.

A first application to quantum annealing, on a random square lattice, reveals that the new driver Hamiltonian improves upon the usual transverse field, though more systematic studies are required to address conclusively this question.
Indeed it will be interesting to optimize the  interplay of the two parameters $\Gamma(t)$ and $\Lambda(t)$ along the annealing schedule and to perform a size scaling analysis.
Morover it will be important to find  other classes of problem Hamiltonians which can benefit more from this technique.

We notice that the range of applicability of the present algorithm can go beyond SQA, since it would be also possible to explore phase diagrams of XYZ sign-problem free Hamiltonians with transverse field, defined on arbitrary lattices, by performing equilibrium simulations with a converged number of Trotter time steps.

\acknowledgements
 This work has been supported by the Swiss National Science Foundation through the National Competence Center in Research QSIT, by the European Research Council through ERC Advanced Grant SIMCOFE and by ODNI, IARPA via MIT Lincoln Laboratory Air Force Contract No. FA8721-05-C-0002. We acknowledge useful discussions with G. E. Santoro, H. Katzgraber, D. Herr and M. Dolfi.

%%%%%%%%%%%%%%%%%%%%%%%%%%%%%%%%%%%%%%%%%%%%%%%%%%

%\bibliographystyle{apsrev4-1}
%\bibliography{bibQA}

%

\end{document}